\documentclass[a4paper,12pt]{article}

\usepackage[T2A]{fontenc}
\usepackage{amsmath,amssymb}
\usepackage{cite}
\usepackage[linktocpage=true,plainpages=false,pdfpagelabels=false]{hyperref}

\usepackage[utf8]{inputenc}
\usepackage[english]{babel}

\newcommand{\dd}{\partial}
\newcommand{\de}{\delta}
\newcommand{\m}{\mu}
\newcommand{\n}{\nu}
\newcommand{\ls}{\left(}
\newcommand{\rs}{\right)}
\newcommand{\al}{\alpha}
\newcommand{\ff}{\varphi}
\newcommand{\ta}{\tau}
\newcommand{\ti}{\tilde}
\newcommand{\str}[1]{\mathrel{\mathop{\longrightarrow}\limits_{#1}}}
\newcommand{\ka}{\varkappa}
\newcommand{\ga}{\gamma}

\newcommand{\sle}{\Longrightarrow}
\renewcommand{\ln}{\log}

\newcommand{\disn}[2]{$$\displaylines{\refstepcounter{equation}%
            \label{#1}\hskip 1em minus 1em #2\hfilneg}$$}
\newcommand{\nom}{\hfil\hskip 1em minus 1em (\theequation)}

\newcommand{\ns}{\hfill\cr\hfill}

\begin{document}

\title{Analysis of the equations of motion\\ of fictitious matter in embedding theory}

\author{
A.~J.~Ziyatdinov\thanks{E-mail: a.2iyat@yandex.ru},
S.~A.~Paston\thanks{E-mail: pastonsergey@gmail.com}\\
{\it Saint Petersburg State University, Saint Petersburg, Russia;}\\
{\it Petersburg Nuclear Physics Institute of NRC KI, Gatchina, Russia}
}
\date{\vskip 15mm}
\maketitle

\begin{abstract}
Embedding theory is a possible modification of general relativity that provides a framework for explaining the observed effects typically attributed to dark matter. The idea of this modification is to consider our spacetime as a four-dimensional surface in a ten-dimensional flat ambient space. The equations of motion in embedding theory can be reformulated as a set of Einstein equations with the contribution of some additional fictitious matter and of equations describing this matter. We analyze static solutions of these equations, which are reduced to fictitious-matter configurations of the wall, string, and ball types. The string case is ultimately described by the Liouville equation, and the ball case is described by its three dimensional analogue. For the string case, we show that as the density contribution decreases at infinity, all solutions to the Liouville equation are rotationally symmetric. For the case of the ball, we show that under the assumption of spherical symmetry, there exists a unique one-parameter family of solutions that are smooth at the center.
\end{abstract}

\newpage

\section{Introduction}
Einstein's General Relativity (GR) perfectly describes the observed effects related to the gravitational interaction. However, beyond a certain scale, nature requires the existence of additional entities -- namely, dark matter (DM) and dark energy -- which cannot be observed by other means. Assuming the existence of dark energy allows for explaining the accelerating expansion of the Universe. Assuming the existence of DM helps resolve numerous discrepancies between observations and predictions of GR on different scales (see, e.g., \cite{1611.09846}). The existing observations are well described by the $\Lambda$CDM (Lambda cold dark matter) model \cite{gorbrub1}. In the framework of this model, DM can be considered as nonrelativistic dust-like matter generating the same gravitational field as ordinary matter, while any nongravitational interaction between DM and ordinary matter is either absent or undetectably weak. The main known properties of DM following from the available observations can be found in \cite{2005.03520}.

If DM interacts with ordinary matter not only gravitationally but also through other mechanisms, then one can attempt to detect it directly. However, despite numerous hypotheses regarding the nature of DM \cite{1605.04909,2111.00363}, all attempts at direct detection have so far failed \cite{1602.03781}. This suggests that DM (and possibly dark energy as well) is not an independent entity but rather an artifact of how the gravitational interaction is described; that is, from the viewpoint of fundamental theory, DM does not exist. This can occur if the exact theory of gravity is not GR but rather an alternative framework whose equations of motion can be recast into the form of GR equations with an additional contribution from DM. Such alternative theories are usually called \emph{modified theories of gravity}.

The idea of replacing DM with a modification of the laws of gravity was likely first implemented within the framework of modified Newtonian dynamics (MOND) approach \cite{mond} in order to explain galaxy rotation curves -- namely, the dependence of stellar velocity on the distance to the galactic center. In the simplest case of a spherically symmetric mass distribution where a star moves in a circular orbit around the galactic center, we have
 \disn{d2}{
\frac{v^2}{r}=\frac{GM(r)}{r^2}\qquad\sle\qquad v(r)=\sqrt{\frac{GM(r)}{r}},
\nom}
where $M(r)$ is the total matter mass in a ball of the radius $r$. In the GR framework (which reduces to the Newtonian law of gravitation in this case), the quantity $v(r)$ must decrease as $1/\sqrt{r}$ in the vacuum region; however, observations often show $v(r)\approx const$, meaning that $v$ reaches a plateau. To explain this, it is usually assumed that in this region, there exists nonluminous (and hence invisible) DM forming a dark halo around the galaxy. Instead, in the MOND approach, it is assumed that for small values of gravitational acceleration, Newton's law -- which requires the gravitational force to decrease as $1/r^2$ -- no longer holds and is replaced with a slower $1/r$ decay. It is important to note that only observations of DM, which is located in the same place of space as ordinary matter, can be replaced with such a modified theory of gravity.

On scales that are larger than galactic ones, the MOND paradigm no longer works so well \cite{1910.04368}. The Bullet Cluster, in which the center of the DM distribution is separated from that of ordinary matter, is most frequently cited in this regard \cite{bulletcluster}. It is logical to assume that when transitioning to a modified theory of gravity, explaining these observations requires more than a mere modification of the behavior of the Newtonian potential; instead, one must introduce extra gravitational degrees of freedom with respect to GR that describe the dynamics of an independently moving fictitious matter identified with DM. Numerous modified theories of gravity, such as $f(R)$ gravity, scalar-tensor theories, and several other frameworks, possess this property (see \cite{1108.6266}). Another variant, namely, the \emph{mimetic} theory of gravity \cite{mukhanov} (its name reflects the fact that gravitational degrees of freedom ``mimic'' the presence of some fictitious matter), was discussed quite often during the past decade.

A specific class of modified gravity theories arises from a particular change of variables. If the change of variables contains differentiation, then after the change of variables, the set of solutions usually expands, and additional dynamic degrees of freedom can appear \cite{statja60}. An example of such a theory is mimetic gravity \cite{mukhanov}, where the differential change of a variable isolating the conformal mode of the metric is performed in the GR action -- the Einstein-Hilbert action.

Another example utilizing the same action is embedding theory, proposed five decades ago \cite{regge} (also referred to as Regge--Teitelboim gravity or embedding gravity). Unlike the change of variables used in mimetic gravity, that used here for the metric
 \disn{r1}{
g_{\m\n}=(\dd_\m y^a)(\dd_\n y^b)\,\eta_{ab}
\nom}
has a deep geometric meaning: the new independent variable $y^a(x^\m)$ is a function of the embedding of the four-dimensional surface into the ten-dimensional ambient space with a flat metric $\eta_{ab}$ (here and hereafter, $\m,\n,\ldots=0,\ldots,3$, and the indices $a,b,\ldots=0,\ldots,9$ label the components of the Lorentz coordinates of the ambient space); furthermore, relation \eqref{r1} defines the induced metric on the embedded surface. Thus, this modification of gravity is based on a simple assumption: our spacetime is a four-dimensional surface in a flat pseudo-Euclidean ambient space (whereas in the GR framework, we assume that spacetime is an abstract pseudo-Riemannian space). The idea of the approach is obviously suggested by the geometric description of strings, which was reflected in the title of the original paper \cite{regge}: ``General relativity `a la string: a progress report.''

As was mentioned in \cite{pavsic85}, the equations of motion of the embedding theory can be written in the form of the system of equations
 \disn{v1}{
G^{\m\n}=\ka \ls T^{\m\n}+\ta^{\m\n}\rs,
\nom}\vskip -2em
 \disn{v2}{
D_\m\Bigl(\ta^{\m\n}\dd_\n y^a\Bigr)=0,
\nom}\vskip -2em
 \disn{v3}{
\eta_{ab}(\dd_\m y^a)(\dd_\n y^b)=g_{\m\n}.
\nom}
Here, $D_\m$ is the covariant derivative, $T^{\m\n}$ is the energy-momentum tensor of ordinary matter, and the quantities $y^a$ and $\ta^{\m\n}=\ta^{\n\m}$, which describe the fictitious matter of embedding theory (FMET), serve as independent variables alongside the metric $g_{\m\n}$. Furthermore, $\ta^{\m\n}$ acts as the FMET energy-momentum tensor. Equation \eqref{v1} is the Einstein equation, whereas Eqs. \eqref{v2}, \eqref{v3} restrict the behavior of fictitious matter.

One can attempt to identify FMET with DM. To verify this identification, it is necessary to study the FMET properties resulting from the governing equations of motion \eqref{v2}, \eqref{v3}. Details about the embedding theory can be found in \cite{statja77} and the literature cited in it. In the framework of the symmetry of Friedmann models, the FMET properties were discussed in \cite{davids01,statja26}. The general equations of FMET motion arising in the nonrelativistic limit under weak Einstein gravity were obtained in \cite{statja68,statja67}; however, these nonlinear equations are generally very difficult to analyze.

In \cite{statja83}, we investigated a more specific situation: the FMET behavior on sub-cosmological scales and at relatively late stages of the cosmic expansion, when the matter density was no longer excessively high. Possible static configurations were analyzed under the assumption that they arose as a result of its clustering by gravitational forces from the initial, almost homogeneous distribution with a positive density. It turned out that one of three types of clustering can occur in different areas of space: a thick wall, a thick string, and a ball. The first two types of clusters are of interest they can potentially be related (though this issue requires further research) to the observed two-dimensional and one-dimensional cosmic structures -- namely, thin leaf-shaped walls and elongated filaments \cite{cosmicweb}.

The third case appears to be the most interesting, namely, spherically symmetric ball-type clustering. Then FMET has the properties of an isothermal ideal gas, and its density decreases as the distance from the distribution center increases as $1/r^2$ at large distances. Given that $M(r)\sim r$ (in a region devoid of any ordinary matter), in accordance with \eqref{d2}, this yields $v(r)\approx const$ -- the frequently observed plateau typically explained by invoking DM.

In this paper, for the cases of strings and balls, we study the possibility of existence of solutions that are alternative to those found in \cite{statja83} and find sufficient conditions under which precisely the found solutions arise. In Section~\ref{krizl}, for convenience, we briefly formulate the problem statement used in \cite{statja83} and write equations that must be studied. In Section~\ref{strun}, we analyze the equations corresponding to the existence of a string-type solution and show that if only a decrease in the FMET density at infinity is assumed, then the solution is necessarily rotationally symmetric and coincides with the solution found in \cite{statja83}. In Section~\ref{shar}, we analyze the equations corresponding to the presence of a ball-type solution in detail and prove that only the one-parameter family (found in \cite{statja83}) of solutions that are smooth at the center necessarily arises in the case of spherical symmetry.

\section{Static FMET clusters}\label{krizl}
To find static solutions to the equations of FMET motion \eqref{v2}, \eqref{v3}, it is useful to rewrite \eqref{v2} as a system of equations (see, e.g., \cite{statja68}):
 \disn{v4}{
D_\m\ta^{\m\n}=0,
\nom}\vskip -2em
 \disn{v5}{
\ta^{\m\n}b^a_{\m\n}=0,
\nom}
where $b^a_{\m\n}=D_\m\dd_\n y^a$ is the second fundamental form of the four-dimensional surface. The first equation is the standard covariant continuity law for the FMET energy-momentum tensor $\ta^{\m\n}$. The second one comprises 6 equations (because $b^a_{\m\n}$ with respect to superscript $a$ is always transverse to the surface) and typically allows one to express the 6 components of the FMET stress tensor $\ta^{ik}$ (here and hereafter, $i,k,\ldots=1,2,3$) in terms of its energy density $\rho_\ta\equiv\ta^{00}$ and momentum density $\ta^{0k}$.

We seek solutions to these equations for which FMET is similar to present-day cold DM, that is, $\ta^{\m\n}\approx\rho_\ta\de^\m_0\de^\n_0$, with $\ta^{0k},\ta^{ik}\ll \rho_\ta$. We assume that the almost homogeneous FMET distribution with $\rho_\ta>0$ appears as a result of the cosmic expansion. As a result of the subsequent growth of density fluctuations, static clusters with the desired structures may emerge. In this asymptotic regime, we neglect regions of strong gravitational field and the effects of cosmic expansion; this implies that the metric $g_{\m\n}$ is close to
 \disn{k4}{
g_{\m\n}=\eta_{\m\n}+h_{\m\n}
\nom}
with a small perturbation $h_{\m\n}$. As a consequence, the embedding function can be written in the form $y^a= \bar y^a+q^a$, where $\bar y^a$ is the background embedding of the metric $\eta_{\m\n}$, and $q^a$ is its small deformation.

If we restrict our consideration to a sufficiently small spacetime region in which the second fundamental form $\bar b^a_{\m\n}$ corresponding to $\bar y^a$ varies slowly, we can choose the coordinate axes in the ambient space so that
 \disn{k11}{
\bar b^a_{\m\n}=\de^a_A B^A_{\m\n},
\nom}
holds approximately; here, $A=4,\ldots,9$, and the quantity $B^A_{\m\n}=B^A_{\n\m}$ is independent of the point, i.e., is a set of constants. This quantity has the dimension of inverse length; we introduce a characteristic length scale $L$ such that $B^A_{\m\n}\sim 1/L$. For self-consistency, the spacetime region under consideration must be smaller than $L$.

Assuming the gravitational field $h_{\m\n}$ (and consequently $q^a$) and its derivatives sufficiently small, we have
 \disn{k13}{
b^a_{\m\n}\approx\bar b^a_{\m\n}.
\nom}
Then, taking expression \eqref{k11} into account, Eq.~\eqref{v5} reduces to
 \disn{k14}{
\ta^{\m\n}B^A_{\m\n}=0
\nom}
in the leading-order approximation within the region under study, and we thus say that FMET is in the linear regime because this equation is linear in $\ta^{\m\n}$. In this case, we have
 \disn{k14.1}{
\ta^{ik}=-\al^{ik}_A\ls \ta^{00}B_{00}^A+2\ta^{0m}B_{0m}^A\rs\approx
-\al^{ik}_A  B_{00}^A \rho_\ta=w^{ik}\rho_\ta,
\nom}
where
 \disn{k15}{
\al^{ik}_A=\al^{ki}_A,\quad
\al^{ik}_A B_{ik}^B=\de^B_A,\qquad
w^{ik}=-B_{00}^A\al^{ik}_A\ll 1.
\nom}
The quantity $w^{ik}$ appearing here varies slowly within the spacetime region under consideration. Consequently, in the linear regime, FMET has a stress tensor that is proportional to its density via this slowly varying matrix coefficient; i.e., FMET behaves as a medium with anisotropic pressure and a linear equation of state.

As the total matter density increases and the gravitational field strengthens, condition \eqref{k13} is violated. Consequently, instead of Eq.~\eqref{k14}, one must invoke the initial equation \eqref{v5}, meaning that FMET ceases to operate in the linear regime. In this regime, the matrix coefficient $w^{ik}$ becomes time-dependent, which prevents the formation of static clusters beyond the FMET linear regime -- specifically, at densities exceeding a certain critical value (see \cite{statja83} for details). This FMET property is very important in the case where we try to see DM in it. If FMET could clump arbitrarily strongly, the question would arise as to why, in addition to clusters corresponding to galaxies, we do not observe much tighter clusters with densities and scales close to those of stars and planets. The maximum possible FMET density depends on $L$ defined by the background embedding $\bar y^a(x^\m)$. To match observational data, one must set $L=4\,\mathrm{Mpc}$ \cite{statja83} (though this scale may vary by an order of magnitude).

In the linear regime of FMET, given the static solution from Eq.~\eqref{v4} and taking Eq.~\eqref{k14.1} into account, the leading-order approximation yields the equation
 \disn{k39}{
\rho_\ta\dd_i\ff+w^{ik}\dd_k\rho_\ta=0.
\nom}
Here, we recall that the solution to the linearized Einstein equations in harmonic coordinates yields $h_{\m\n}=2\ff\de_{\m\n}$, where $\ff$ is the Newtonian gravitational potential satisfying the Poisson equation
 \disn{k47.0}{
\dd_i\dd_i\ff=4\pi G \rho_\ta,
\nom}
where $G$ is the gravitational constant, if there is no ordinary matter. Omitting the trivial solution $\rho_\ta=0$, Eq.~\eqref{k39} can be rewritten in the form
 \disn{k40}{
\dd_i\ff+w^{ik}\dd_k z=0,
\nom}
where $z=\ln(\rho_\ta/\tilde\rho)$, and $\tilde\rho$ is an arbitrary positive dimension-making constant. Now, \eqref{k47.0} can be rewritten as
 \disn{k47}{
\dd_i\dd_i\ff=4\pi G \tilde\rho\, e^z
\nom}
and the problem of seeking static FMET clusters reduces to solving the system of equations \eqref{k40}, \eqref{k47}.

By diagonalizing the matrix $w^{ik}$ and analyzing different cases of coincidence/noncoincidence of its eigenvalues, we show that the solution to Eq.~\eqref{k40} is always given by $z=C_1-\ff/w$, where $w\ne0$ is any of the eigenvalues and $C_1$ is an integration constant. In this framework, if all eigenvalues are distinct, then for $w>0$, static clusters with densities depending on a single coordinate -- namely, thick-wall clusters -- can emerge. If two of the three eigenvalues coincide, the density depends on two coordinates, yielding thick-string clusters. Finally, if all three eigenvalues coincide, the density depends on all three coordinates, giving rise to spherically symmetric ball-type clusters \cite{statja83}. The longitudinal dimensions of the wall and the string are determined by the dimension of the area where the underlying assumptions remain true; they must be smaller than $L$.

In all the three cases, Eq.~\eqref{k47} transforms into the equation
 \disn{k47.1}{
\Delta\ff=C_2 e^{-\ff/w},
\nom}
where $C_2$ is a positive constant, and $\Delta$ is the Laplace operator in one, two, or three dimensions, respectively. If the independent variable $x=\ti x\sqrt{w/C_2}$ (the variable $\ti x$ is dimensionless) and the unknown function $\ff=-w\ti\ff$ are changed, then the equation acquires the form
 \disn{k47.2}{
\Delta\ti\ff(\ti x)+e^{\ti\ff(\ti x)}=0.
\nom}
which contains no parameters. In the one-dimensional case, such an equation is easily solved. In the two-dimensional case, this equation is well known as the Liouville equation \cite{Liouville1853}; we analyze its solutions in Section~\ref{strun}. Section~\ref{shar} is devoted to discussing three-dimensional case.

\section{Solution of the thick-string type}\label{strun}
We now consider the two-dimensional case of Eq.~\eqref{k47.2}, where $\Delta$ is the two-dimensional Laplace operator. This relation is known as the Liouville equation \cite{Liouville1853}; it has been extensively studied. In particular, it was proved in \cite{chen-li} that if the integral
\begin{equation}\label{am1}
\int\! d^2\ti x\,\, e^{\ti\ff(\ti x)}
\end{equation}
is finite, which, from a physical perspective, corresponds to a finite total mass of the static configuration under consideration in the two-dimensional case, then all solutions to Eq.~\eqref{k47.2} are necessarily symmetric with respect to the rotations about a certain center. All such solutions can easily be found as solutions of the problem
 \disn{k62}{
\ti\ff''(s)+\frac{1}{s}\ti\ff'(s)+e^{\ti\ff(s)}=0,\qquad
s\ti\ff'(s)\str{s\to0}0,
\nom}
$s=\sqrt{(\ti x_1-\hat x_1)^2+(\ti x_2-\hat x_2)^2}$ is the two-dimensional distance to an arbitrarily chosen center $\hat x$; here and hereafter, the prime denotes differentiation with respect to the argument of the function. This solution has the form
 \disn{k67}{
\ti\ff(s)=\ti\ff_0-2\ln\ls 1+\frac{s^2}{8} e^{\ti\ff_0}\rs,
\nom}
where $\ti\ff_0$ is the integration constant. The FMET cluster with the density profile of the form
 \disn{k69}{
\rho_\ta=\frac{\rho_0}{\ls 1+\dfrac{s^2}{\de^2}\rs^2},
\nom}
corresponds to it; here, $\rho_0=2w/(\pi G\de^2)$ and $\de$ is the constant parameterizing the solution \cite{statja83}.

At first glance, the condition of a finite total mass, i.e., the convergence of the integral \eqref{am1}, seems to be physically necessary. However, for example, in the three-dimensional case of the spherically symmetric DM distribution, to implement the observed plateauing of the rotation curve $v(r)$, as can be seen from \eqref{d2}, the behavior of $M(r)\sim r$ is necessary for large $r$, which corresponds to the decrease in the matter density as $1/r^2$. Such a slow decrease does not correspond to the finite total mass; therefore, we have to assume that this behavior is true only in some finite area of the radius variation. In the case of string-type clusters, this consideration also forces one to consider what the solutions might be if the density decreases more slowly than is required for the finiteness of integral \eqref{am1}.

This question is answered in the present section. We show that the aforementioned statement from \cite{chen-li} can be strengthened as follows: for a solution to the two-dimensional equation \eqref{k47.2} to be necessarily symmetric with respect to rotations about a certain center (meaning that it takes the form \eqref{k67}), it is sufficient to require only the decrease in the quantity $e^{\ti\ff(\ti x)}$, which corresponds to the density, for large $\ti x$.

To prove this assertion, we use the general solution of the Liouville equation \eqref{k47.2}, which is known (see, e.g., \cite{Henrici1993}) and is written in the form
\begin{equation}\label{eq4}
\ti\varphi(\ti x)=\ln{\frac{|f'(z)|^2}{\ls 1+\frac{1}{8}|f(z)|^2\rs^2}},
\end{equation}
where $z=\ti x_1+i\ti x_2$ and $f(z)$ is a meromorphic function whose arbitrary choice is only restricted by the condition that its derivative $f'(z)$ vanishes nowhere. To understand how the conditions for the function $\ti\varphi(\ti x)$ given in the assertion impose the constraint on the function $f(z)$, we consider two possible variants of the behavior of $|f(z)|$ as $z\to\infty$:
\begin{enumerate}
\item $|f(z)|\longrightarrow\infty$,
\item $|f(z)|<C$, where $C$ is a constant.
\end{enumerate}

We begin the analysis with the first variant $|f(z)|\to\infty$. For this case, taking into account that according to the condition of the proved assertion, $e^{\ti\ff}$ tends to zero as $z\to\infty$ in this limit, from \eqref{eq4}, we have the asymptotics
\begin{equation}\label{eq5}
e^{\ti\ff}=\frac{|f'(z)|^2}{\ls 1+\frac{1}{8}|f(z)|^2\rs^2}\approx 64 \left|\frac{f'(z)}{f(z)^2}\right|^2=
64 \left|\left(\frac{1}{f(z)}\right)'\right|^2\qquad\sle\qquad
\left(\frac{1}{f(z)}\right)'\str{z\to\infty} 0.
\end{equation}
Since $f(z)$ is a meromorphic function, $1/f(z)$ and $\ls 1/f(z)\rs'$ are also meromorphic. As a result, for large $z$, as a consequence of Eq.~\eqref{eq5}, the asymptotic relation
\begin{equation}\label{eq66}
\left(\frac{1}{f(z)}\right)'\approx\frac{C_1}{z^{n}},
\end{equation}
holds. Here, $n\in\mathbb{N}$, and $C_1,C_2,\ldots$ are constants here and hereafter. Because $n=1$ contradicts the meromorphic property of $1/f(z)$, we obtain the constraint $n\ge 2$ additionally.

From \eqref{eq66}, using the integration, we obtain the asymptotics as $z\to\infty$ in the form
\begin{equation}
\label{eq6}
\frac{1}{f(z)}\approx C_2-\frac{C_1}{(n-1)z^{n-1}}.
\end{equation}
In view of the fact that in the case of $|f(z)|\to\infty$ under consideration, it can be concluded that $C_2=0$, which means that as $z\to\infty$, we have
\disn{sp1}{
\frac{1}{f(z)}\approx -\frac{C_1}{(n-1)z^{n-1}}\qquad\sle\qquad
f(z)\approx -\frac{n-1}{C_1}z^{n-1}\qquad\sle\ns\sle\qquad
f'(z)\approx -\frac{(n-1)^2}{C_1}z^{n-2}\qquad\sle\qquad
\frac{1}{f'(z)}\approx -\frac{C_1}{(n-1)^2 z^{n-2}}.
\nom}
Since $f'(z)$ is a meromorphic function that vanishes nowhere (see the comments following Eq.~\eqref{eq4}), it follows that $1/f'(z)$ is an entire function. In this case, since $n\ge 2$, relation \eqref{sp1} further implies that it is bounded. Consequently, by Liouville's theorem, this function is constant, which means that $f(z)=C_3(z-z_0)$. Using this in \eqref{eq4}, we obtain the solution in the form
\begin{equation}\label{sp2}
\ti\varphi=\ln{\frac{|C_3|^2}{(1+\frac{1}{8}|C_3|^2|z-z_0|^2)^2}}
\end{equation}
and, in terms of real variables $\ti x$, it obviously turns out to be symmetric with respect to rotations about the center $\hat x$ corresponding to the point $z_0$, which corresponds to the proved assertion. It is also easy to note that obtained solution \eqref{sp2} reproduces Eq.~\eqref{k67} up to the notation of the constants.

Next, we consider the second variant among the three ones, namely, the case $|f(z)|<C$ as $z\to\infty$. In this case, due to the assumed vanishing of the quantity $e^{\ti\ff}$ as $z\to\infty$, one can conclude from \eqref{eq4} that $f'(z)\to0$ in this limit. As a result, taking into account that the function $f'(z)$ is meromorphic, for large $z$, the asymptotics
\disn{sp2.1}{
f'(z)\approx\frac{C_1}{z^{n}},
\nom}
holds, where $n\in\mathbb{N}$, and we again conclude that $n\ge 2$ because $f(z)$ is meromorphic. From \eqref{sp2.1}, in particular, it follows that the function $1/f'(z)$ is polynomially restricted as $z\to\infty$. On the other hand, as previously noted, $1/f'(z)$ is an entire function, which, by the generalized Liouville theorem, implies that $1/f'(z)$ is a polynomial $P(z)$. Moreover, it is easy to see that its degree must be $n$, and hence it is at least 2.

We find that the derivative $f'(z)$ of the meromorphic function $f(z)$ is $1/P(z)$. Denoting the roots of the polynomial $P(z)$ by $z_j$ and their respective multiplicities by $m_j$, we can write
\disn{sp3}{
f'(z)=C_1\prod_{j=1}^k \frac{1}{(z-z_j)^{m_j}}.
\nom}
The function $f(z)$ must have poles only at the same points $z_j$; furthermore, their multiplicities must be lower by one, equaling $m_j-1$ (which means that $m_j\ge 2$ in accordance with the meromorphic nature of $f(z)$). Therefore, it can be written as
\disn{sp4}{
f(z)=Q(z)\prod_{j=1}^k \frac{1}{(z-z_j)^{m_j-1}},
\nom}
where $Q(z)$ is a polynomial; we let $l$ denote its degree. We note that as $z\to\infty$, accordance to Eq.~\eqref{sp4}, the function $f(z)$ scales asymptotically as $z$ to the power of $l-\sum_{j=1}^k(m_j-1)$, whereas according to \eqref{sp3}, $f'(z)$ scales asymptotically as $z$ to the power of $-\sum_{j=1}^k m_j$. On the other hand, since these powers must differ by one, we obtain
\disn{sp5}{
-\sum_{j=1}^k m_j=\ls l-\sum_{j=1}^k(m_j-1)\rs-1\qquad\sle\qquad k+l=1.
\nom}

Since $k$ is the number of distinct roots of the polynomial $P(z)$ (which is a nonnegative integer) and $l\ge0$, we find that only $k=0$ or $k=1$ is possible. The first possibility implies that $f'(z)= const$, which is impossible under the considered case of $|f(z)|<C$ as $z\to\infty$. Consequently, only the possibility $k=1$ remains (in this case, $l=0$), meaning that the polynomial $P(z)$ has only one root. Its multiplicity $m$ must be at least 2 because, as was mentioned above, the degree of the polynomial $P(z)$ is at least 2. As a result, in accordance with \eqref{sp4}, we have
\disn{sp6}{
f(z)=\frac{C_2}{(z-z_1)^{m-1}}.
\nom}
\begin{equation}\label{sp7}
\ti\varphi=\ln{\frac{(m-1)^2|C_2|^2}{(|z-z_1|^m+\frac{1}{8}|C_2|^2|z-z_1|^{2-m})^2}}.
\end{equation}
For the resulting solution, the rotational symmetry in terms of the real variables $\ti x$ about the center $\hat x$ corresponding to the point $z_1$ is immediately apparent, which confirms the proved assertion. Thus, the proof is complete. In addition, we note that solution \eqref{sp7} coincides with solution \eqref{sp2} for $m=2$ (assuming $C_2=8/C_3$), despite being obtained from another function $f(z)$. It can also be noted that for other values of $m$, rotationally symmetric solution \eqref{sp7} has an asymptotics at $z=0$ such that it corresponds to the solution of Eq.~\eqref{k62}, but does not correspond to the required asymptotics as $s\to0$.

Thus, it can be concluded that if, for large $\ti x$, we require only the decrease in the quantity $e^{\ti\ff(\ti x)}$, which corresponds to the FMET density $\rho_\ta$, then all solutions of the two-dimensional Liouville equation \eqref{k47.2} turn out to be symmetric with respect to rotations about a certain center, which means that they are the solution \eqref{k67}, \eqref{k69} used in \cite{statja83}.

\section{Solution of the ball type}\label{shar}
We now consider the three-dimensional case of Eq.~\eqref{k47.2}, where $\Delta$ is the three-dimensional Laplace operator. As in the two-dimensional case, it is logical to assume that the quantity $e^{\ti\ff(\ti x)}$, which corresponds to the FMET density, decreases for large $\ti x$. Unfortunately, in the three-dimensional case, there is no known proof of spherical symmetry for all solutions under this assumption or even under stronger ones -- such as the finiteness of the total mass. Therefore, we restrict ourselves to studying spherically symmetric solutions; unexpectedly, this problem turns out to be noticeably more complex than the two-dimensional case. In this case, there is no need to require the decrease in the density $e^{\ti\ff(\ti x)}$ for large $\ti x$ except the substantiation of the choice of $w>0$ in Eq.~\eqref{k47.1}.

The search for spherically symmetric solutions of the three-dimensional equation \eqref{k47.2} reduces to solving the problem
\begin{equation}\label{eq5.2}
\ti\varphi''(r)+\frac{2}{r}\ti\varphi'(r)+e^{\ti\ff(r)}=0,
\end{equation}
\begin{equation}\label{eq6.2}
r^2\ti\varphi'(r)\str{r\rightarrow0}0,
\end{equation}
$r=\sqrt{\ti x_1^2+\ti x_2^2+\ti x_3^2}$ is the radius. We note that condition \eqref{eq6.2}  arises from the requirement that the
quantity $e^{\tilde{\varphi}(r)}$, which corresponds to the density, lacks a delta-function contribution at $r=0$.

We now analyze \eqref{eq5.2}. As shown in \cite{statja83}, under the change of variables $r=e^{u}$ and $\tilde{\varphi}(r)=\gamma(u)-2u$, this equation reduces to
 \disn{k80}{
\ga''(u)+\ga'(u)-2+e^{\ga(u)}=0.
\nom}
Since this equation does not explicitly contains the independent variable $u$, its order can be reduced via the substitution $\psi(\ga)=\ga'(u)$, yielding the equation
 \disn{k82}{
\psi(\ga)\psi'(\ga)+\psi(\ga)-2+e^{\ga}=0.
\nom}
As noted in \cite{statja83}, the transition from \eqref{k80} to \eqref{k82} results in the loss of the constant solution to Eq.~\eqref{k80} of the form $\ga(u)=\ln2$. It corresponds to the solution
 \disn{k82.1}{
\ti\ff(r)=\ln\frac{2}{r^2}
\nom}
(for which condition \eqref{eq6.2} is easily verified), yielding a density distribution profile of the form
 \disn{k83}{
\rho_\ta=\frac{w}{2\pi G\, r^2},
\nom}
This configuration, known as the singular isothermal sphere \cite{GalDyn}, diverges at the center. We seek other solutions to Eq.~\eqref{k80} determined by the solutions to Eq.~ \eqref{k82} that satisfy condition \eqref{eq6.2} as $r\to0$. This order-reducing change of variables corresponds to the relation
\begin{equation}\label{int}
\int\frac{d\ga}{\psi(\ga)}=u+C_1,
\end{equation}
where an indefinite integral is assumed. Taking into account that $u\to-\infty$ as $r\to0$, we conclude that the integral must diverge as $\ga$ approaches the value corresponding to the $r\to0$ limit. There are three possible behaviors for $\ga$ under which this divergence can occur:
\begin{enumerate}
\item $\ga\to\infty$;
\item $\ga\to\ga_0$, where $\ga_0$ is a finite value;
\item $\ga\to-\infty$.
\end{enumerate}

We consider these cases in turn.

First, to analyze case 1, we introduce the change of variables $v=e^\ga$ and $s(v)=(\psi(\ga))^2$ into \eqref{k82}, which yields $s'(v)$
\begin{equation}\label{eq20.2}
\frac{s'(v)}{2}+1=\frac{2\pm\sqrt{s(v)}}{v}.
\end{equation}
In this case, as $r\to0$, we must have $\ga \to\infty$, implying that $v\to\infty$. We now examine the behavior of the right-hand side of Eq.~\eqref{eq20.2} in this case. If it vanishes in this limit, then according to the equation, the solution must behave asymptotically as $s(v)\approx C_2-2v$ as $v\to\infty$; however, this contradicts the requirement that $s(v)$ be strictly positive. If the right-hand side tends to a nonzero finite value as $v\to\infty$, then one must have $\sqrt{s(v)}\sim v$, implying that $s'(v)\sim v$. This, in turn, means that the left-hand side of the equation diverges, leading to a contradiction. If the right-hand side of \eqref{eq20.2} diverges (with either sign) as $v\to\infty$, then the term 1 on the left-hand side and the term 2 in the numerator of the right-hand side can be neglected in this asymptotic regime. This simplifies the equation, making it straightforward to find the asymptotic solution $s(v)\sim(\ln v)^2$, which contradicts the assumed growth of the right-hand side. Consequently, case 1 necessarily leads to a contradiction, implying that no solution to Eq.~\eqref{k82} exists with the asymptotics $\ga\to\infty$.

Now we study case 2, where the integral in \eqref{int} diverges as $\ga=\ga_0$. For this to occur, the condition $\psi(\ga_0)=0$ must hold. We rewrite Eq.~\eqref{k82} in the form
\begin{equation}\label{eq24}
\frac{d\psi}{d\ga}+1=-\frac{2}{\psi}(e^{\ga-\ln{2}}-1).
\end{equation}
and consider it in the neighborhood of the point $\ga=\ga_0$ where $\psi=0$. If $\ga_0\ne\ln2$, then the right-hand side grows infinitely as this point is approached, and the term 1 on the left-hand side can be neglected in comparison. Within the framework of studying the leading asymptotics, replacing $\ga$ with $\ga_0$ on the right-hand side yields a solution of the form $\psi(\ga)\sim\sqrt{\ga-\ga_0}$. As is easy to see, it does not lead to the divergence of the integral in \eqref{int} at $\ga=\ga_0$, which results in a contradiction; therefore, $\ga_0=\ln2$. Then, as $\ga\to\ga_0$, $\psi\to0$, the right-hand side of \eqref{eq24} has the asymptotics $-2(\ga-\ga_0)/\psi$. In the limit under consideration, the derivative $d\psi/d\ga$ in the left-hand side of \eqref{eq24} can be written as the asymptotics of the quantity $\psi/(\ga-\ga_0)$. If $(\ga-\ga_0)/\psi$ is formally denoted as $\ga'(0)$ (it is not important whether the value of the derivative $\ga'(\psi)$, which is the limit, exists at $\psi=0$ and whether it is finite), the asymptotics of Eq.~\eqref{eq24} can be written as
 \disn{sp10}{
\frac{1}{\ga'(0)}+1=-2\ga'(0)\qquad\sle\qquad
2\ga'(0)^2+\ga'(0)+1=0.
\nom}
as $\psi\to0$ and $\ga\to\ga_0$. Since this quadratic equation has no real roots, it follows that no solutions to Eq.~\eqref{k82} exist with the asymptotics $\ga\to\ga_0$ corresponding to case 2.

Finally, we examine case 3, where the integral in \eqref{int} diverges as $\ga \to-\infty$. For large negative $\ga$, the contribution $e^\ga$ can be neglected in Eq.~\eqref{k82}; therefore, in this regime, the equation
 \disn{sp11}{
\psi(\ga)\psi'(\ga)+\psi(\ga)-2=0.
\nom}
holds. Except for the special solution $\psi(\ga)=2$, all of its solutions can be written in terms of an inverse function of the form:
 \disn{sp12}{
\ga(\psi)=C_1-\psi-2\ln|\psi-2|.
\nom}

Consequently, for the leading-order asymptotics of $\psi(\ga)$ at large negative $\ga$, we find $\psi(\ga)=-\ga$. Substituting this into \eqref{int} yields $\ga(u)\sim-e^{-u}$ as $u \to-\infty$ (which, it is worth recalling, corresponds to $r\to0$; see the discussion following Eq.~\eqref{int}). If condition \eqref{eq6.2} in this limit is rewritten in terms of $\ga(u)$, it takes the form
 \disn{sp13}{
e^{u}\big(\ga'(u)-u\big)\str{u\to-\infty}0,
\nom}

It is easy to see that the obtained asymptotics $\ga(u)\sim-e^{-u}$ contradicts this condition. This implies that we fail to find solutions to Eq.~\eqref{k82} that satisfy condition \eqref{eq6.2} within the class of functions with the asymptotics $\ga \to-\infty$ (corresponding to case 3), with the sole exception of the case where $\psi(\ga)=2$ is chosen as the solution to Eq.~\eqref{sp11}. Assuming this single remaining case -- where the leading-order asymptotics of $\psi(\ga)$ takes the form $\psi(\ga)=2$ for large negative $\ga$ -- and substituting it into Eq.~\eqref{int}, we obtain $\ga(u)\sim2u$ as $u \to-\infty$. It is easy to see that condition \eqref{sp13} is satisfied for this asymptotic behavior. Consequently, we find that, for problem \eqref{eq5.2}, \eqref{eq6.2}, there exist only two types of solutions: the aforementioned solution \eqref{k82.1}, which is singular at $r=0$, and the solution arising from Eq.~\eqref{k82} with the boundary condition
 \disn{sp14}{
\psi(\ga)\str{\ga\to-\infty}2.
\nom}

This last solution determines the ball-type spherically symmetric FMET clustering found in \cite{statja83}. The corresponding function $\ti\ff(r)$ remains finite as $r\to0$. This function also appears in stellar dynamics when describing a self-gravitating isothermal sphere \cite{GalDyn} (Sec. 4.3.3(b)). The transition from the function $\psi(\ga)$ to $\ga(u)$ via relation \eqref{int} is ambiguous due to the presence of an arbitrary integration constant $C_1$ in Eq.~\eqref{int} (though the subsequent transition to $\ti\ff(r)$ is unique). The choice of $C_1$ ensures the fulfillment of the additional requirement $\ti\ff(0)=0$, which makes the function $\ti\ff(r)$ unique. This unique function can be used as a special function $\hat\ff(r)$ (which cannot be expressed in terms of known special functions) evaluated numerically. It is smooth for $r\ge0$, and its plot was provided in \cite{statja83}. One can then write the general solution corresponding to an arbitrary value of the constant $C_1$ in terms of the introduced special function:
 \disn{k88}{
\ti\ff(r)=\hat\ff(r e^{C_1})+2C_1.
\nom}
The FMET clustering corresponding to this solution has the density
 \disn{k89}{
\rho_\ta=\rho_0 \exp{\ls\hat\ff\ls\frac{4r}{\de}\rs\rs},
\nom}
where $\rho_0=4w/(\pi G\de^2)$ and $\de$ is the constant parameterizing the solution \cite{statja83}. Thus, we have shown that this solution necessarily emerges as the unique smooth solution at $r=0$ under the assumption of spherical symmetry.

\textbf{Acknowledgments.}
The authors are grateful to V.V.~Sukhanov for the useful discussions. The authors also thank the organizers of the VIII International Conference ``Models of quantum field theory'' (MQFT-2025), which was dedicated to Professor Alexander Nikolaevich Vasil'ev.

\textbf{Funding.} This work was supported by the Ministry of Science and Higher Education of the Russian Federation (agreement 075-15-2025-343 dated 29/04/2025 for Saint Petersburg Leonhard Euler International Mathematical Institute at Saint Petersburg State University).

\textbf{Conflict of interest.} The authors of this work declare that they have no conflicts of interest.


\end{document}